# Microscopic origin of collective exponentially small resistance states

J. C. Phillips

*Dept. of Physics and Astronomy, Rutgers University, Piscataway, N. J., 08854-8019*

## Abstract

The formation of "zero" (exponentially small) resistance states (ESRS) in high mobility two-dimensional electron systems (2DES) in a static magnetic field B and subjected to strong microwave (MW) radiation has attracted great theoretical interest. These states appear to be associated with a new kind of energy gap _. Here I show that the energy gap _ is explained by a microscopic quantum model that involves the Prime Number Theorem, hitherto reserved for only mathematical contexts. The model also contains the zeroes of the zeta function, and explains the physical origin of the Riemann hypothesis.

Two experimental groups have discovered giant magnetoresistance oscillations associated with low-field cyclotron resonance in samples previously used to study quantum Hall effects. In very high mobility samples the minima of these oscillations correspond to exponentially small resistance[1-3]. There are many suggested analogies to explain this entirely unexpected phenomenon[1-9]. Photon-assisted formation of collective tunneling states between initial and final states involving multiquantum well (MQW) geometries is well known[10]. Theories predicted[11] dynamic localization and absolute negative conductance due to coherent tunneling in semiconductor superlattices subjected to ac electric fields nearly 20 years before a similar effect associated with sequential incoherent tunneling was observed[12]. Many examples of this phenomenon are now available, all involving conductance oscillations generated by oscillations in the density of states. Thus when oscillations in the resistance of a 2DES in a static magnetic field B and subjected to strong microwave (MW) radiation were first observed[1], it seemed natural to many theorists[5-9] to construct models that attributed these oscillations to the oscillatory density of states (ODOS) associated with Landau levels. ODOS models are characteristically vague, however, on the question of how one defines *separate* densities of initial and final states in a *structureless* 2DES that are consistent with uniform phase-space requirements. Moreover, the observed resistivities are ohmic, or nearly so, in strong contrast to the extremely non-linear I-V characteristics observed with transitions between initial and final states in tunnel junctions[12].

When the mobility _ of the 2DES was increased by a factor of 5, an even more mysterious effect appeared[2]. At higher power levels near Landau level filling of j = 1/4, the MW-induced oscillatory negative resistance oscillation *cancels* the background resistance, leaving an exponentially small resistance. Several experiments with different sample geometries have shown[13] that this cancellation is not a boundary or edge effect, but it is associated with the formation of some kind of bulk collective state with an energy gap. It appears that such an energy gap is not obtainable from ODOS models.

When the mobility _ of the 2DES was further increased by another factor of 1.6, more detail appeared in the I-V characteristic[3]. The frequency _ dependence (at constant power P) of the oscillatory maxima and minima in the resistance can be seen to be qualitatively different, for instance, in Fig. 2 of ref. 3. The maxima associated with integer filling sharpen with increasing _, as one would expect from cyclotron resonance with increasing __, but the minima are characterized by activation energies $T_0$ that are less sensitive to frequency. However, $T_0$ is very sensitive to _: the increase of _ by a factor of 1.6 increased $T_0$ from 10K to 20K! These features are certainly inconsistent with ODOS models, and they indicate that there is a fundamental, not merely oscillatory, dichotomy between the mechanisms responsible for the maxima and minima in the resistance.

Topology suggests[14] a natural origin for this dichotomy in terms of closed cyclotron orbits _ (which obviously produce magnetoresistance maxima) and open orbits _ (which are thought to produce negative magnetoresistance, for example, along misfit dislocation lines in 2DES[15]). It has been argued[14] semiclassically that even in a structureless 2DES _ orbits can be separated from _ orbits by drawing on the MW magnetic field gradient fluctuations[16] to exploit the excess configurational entropy[2] that reaches its maxima at j = 1/4. The open orbits _ are favored because they screen internal Coulomb fields better than closed orbits _.

While such semiclassical reasoning is most plausible, it does not provide a fully satisfactory explanation for ESRS. There are two ways that the semiclassical mechanism could fail. First, the Coulomb central forces could be so strong that they actually destroy open orbits by occupying some closed orbits (configuration space rigid). If that does not happen, the alternative is that there are too many ways to form open orbits, and that only some of them are utilized (configuration space floppy). In that case, alternative open orbits would exist, and carriers in the occupied open orbits could be scattered by phonons[17] into unoccupied open orbits, giving rise to some non-exponential background resistivity in a sample with arbitrarily large (but not infinite) mobility. In other words, there is a problem in counting quantum states in this strongly disordered medium where **k** is not a good quantum number, even in the absence of external fields[18].

The problem of matching numbers of forces (or Lagrangian constraints $N_c$) to degrees of freedom $N_d$ has previously arisen in other contexts, notably network glasses, where it leads[19] to the limbo condition for the ideal (strain-free) glass in a mean-field approximation,

$$N_c = N_d \qquad (1)$$

Recent results show that in the presence of local field corrections this exact condition broadens into a robust intermediate topological phase where there are two stiffness transitions and a reversibility window[20-22]. However, counting states and forces belonging to different spaces in a naturally hierarchical molecular glass appears to be much easier than counting them in a 2DES. Is there some way of counting open orbits in an apparently continuous 2DES?

There is. By definition an open orbit is a coherent path. If a carrier is scattered at any point by the combined internal field fluctuations (including phonons), a new path (which may be open or closed) begins. We can imagine laying down a mesh of arbitrary fineness centered on the starting point of either a one-electron or many-electron orbit. We now wind outward from the center of the mesh, labeling points with integers. An interrupted or segmental path can be described by a composite product of such integers. For an open orbit to be optimal (coherent) it should consist of just one segment, in other words, it should be labeled by an integer with no factors, a prime number. *All coherent open orbits can be topologically labeled by prime numbers.*

Pursuing the analogy with an ideal glassy network, our next task is to understand the hierarchy of many-electron forces. Because of the finite thickness (*quasi*-two-dimensional) nature of depletion layers, electron-electron quasi-particle Coulomb interactions consist of two parts, $V_2 \sim r^{-1}$ for large r, and $V_1 \sim \ln r$ for small r. The relative weights of $V_1$ and $V_2$ in determining collective many-electron wave functions are not known. However, it has been observed[23] that the neglect of $V_1$ predicts an excitation energy for collective $\_ = 1/3$ fractional quantum Hall states that is too large by a factor of 4, and a $B^{1/2}$ scaling of this energy, based on there being only one length scale (the magnetic length), whereas observed activation energies depend only weakly on B. This is similar to the weaker frequency dependence observed for ESRS, and it suggests that the dynamics of such states are dominated by $V_1$.

To complete the analogy with an ideal glassy network, we must find a relation analogous to (1), which can be rewritten as $(N_d)^{-1} N_c = 1$. The suggested analogy is

$$N(\_) V_1 = O(1) \qquad (2)$$

where by O(1) is meant a number of order unity that is independent of r, and N(\_) is the density of open orbit \_ states. The number of such states depends on the number of primes up to p, and the latter is proportional to $N(p)p \sim N(p)r^d$. The density $N(p)$ of prime numbers is given by the prime number theorem (PNT). PNT, as conjectured by Legendre and Gauss (1800), and proved a century later, states[24] that

$$N(p) \sim (\ln p)^{-1} \qquad (3)$$

Thus (2) is satisfied, and the new collective state responsible for ESRS is an ideal glass. Window glass is another, somewhat more familiar, ideal glass[25].

The scattering rate for a set of N open orbits $\{\_\}$ into another set $\{\_\_\}$ is reduced relative to all final state orbits by an exponentially small factor of order $\exp[-(N/\ln r)]$. The open orbits that produce ESRS are self-organized in the sense that open orbits can optimally screen internal electric fields. A characteristic feature of self-organized systems, such as ideal glasses, is that they occupy an exponentially small fraction of configuration space. Such states can exist as metastable phases sustained by an external supply of free energy (here the low-entropy MW field)[4,14]. So long as the ESRS screens the MW field, it will be stable against thermal fluctuations, as it can scatter into only an equally exponentially small fraction of configuration space. This produces ESRS, and such a state can be regarded as stabilized[2] by an energy gap $\_$. Moreover, when the scattering is dominated by phonons[17], within the adiabatic approximation the carriers screen the phonons, which is reminiscent of superconductivity, but without the Meissner effect. At the same time, the actual geometry of the open orbits will be pinned by ionized impurities within the bulk and at sample surfaces, giving rise to hysteretic effects that resemble pinned or sliding charge density waves[26]. Thus both of the theoretical models proposed to describe experiments[1,2] agree with the more abstract configuration space model discussed here.

The idea that number theory can be related to the properties of Landau levels of a 2DEG will probably seem arcane to most scientists, and should astonish most mathematicians. However, the idea has rich and intriguing possibilities. The Riemann zeta function

$$\zeta(s) = \sum n^{-s} \qquad (4)$$

can be written (Euler, 1737) as

$$\zeta(s) = \prod (1 - p^{-s})^{-1} \qquad (5)$$

where p is prime and $\Gamma(s)$ is the usual factorial function $\Gamma(s) = s\Gamma(s-1)$. Riemann's eight-page paper (1859) emphasized the symmetry properties of $\zeta$ and $\Gamma$ through the functional equation (constant factors of order unity omitted)

$$\Gamma(s/2 - 1)\, \zeta(s) = \Gamma((1-s)/2 - 1)\, \zeta(1-s) \qquad (6)$$

which shows that replacing s by 1-s leaves this product unchanged. Thus s may play the role of energy, with s = 1/2 as the Fermi energy, and a pseudogap in the interval [0,1]. In other words, the algebra of the $\zeta$ function is electron-hole symmetric and it may contain an energy gap. It thus reproduces two of the features identified experimentally[2] from the positions of the resistivity minima at j = 1/4 (not 1/2).

Riemann identified several correction terms to (3). $(\ln p)^{-1}$ "should be considered to be an approximation not to the density of primes as Gauss suggested, but rather to the density of primes plus $\frac{1}{2}$ the density of prime squares, plus, etc."[27] Physically the density of prime squares is suggestive of backscattering (some kind of weak localization), a weak effect that might occur, but for which there is no evidence at present. There are also periodic correction terms, for which there are few definite results.

$\zeta(s)$ may be analytically continued to the entire *s*-plane, except for a simple pole at *s* = 0. The Riemann hypothesis, unquestionably the most celebrated unsolved problem in mathematics, states that the complex zeros of $\zeta(s_1 + is_2)$ all have real part $s_1$ at the "Fermi energy" $s_1$ = 1/2. PNT states that $p_n \sim n\log n$. Now suppose that $p_n$ were *exactly* $n\log n$. In other words, in the Euler product $\zeta$ above, in the spirit of mean-field theory, replace the *n*th prime by *n*log*n*. In this way one defines[28] a pseudo zeta function C(*s*) for Re $s > 1$ that can be analytically continued at least into the half-plane Re $s > 0$. The pseudo zeta function $C(s)$ has no complex zeros whatsoever. This means that the complex zeros of the zeta function are associated with the irregularity of the distribution of the primes (localized states). If one identifies the complex zeros with decaying states scattering from localized states, their mean-field disappearance implies that if instead of (2) one had $N(\_)V_1 = 1$ exactly (in other words, $V = V_1$ everywhere), the resistance minimum could be very deep. This explains how it is possible for resistivity oscillations measured in the MW field at T ~ 1K to improve in quality with sample mobilities _ measured at < 10 mK, even though resistivities measured at 1K without the MW field would be the same.

The identification of the zeroes of _ (s) with localized states suggests a possible physical meaning for _ (s): it could be the participation amplitude $<\_(s)\_k>$ of states in the magnetic and MW fields with all plane wave (extended) states. While this does not bring us any closer to proving Riemann's hypothesis, it does suggest that attempts to do so using field-free inverse potential methods are unpromising. However, there are interesting connections between the asymptotic properties of random matrices and the density of the zeroes of _ (s), and these random matrices exhibit symptoms of a metal-insulator transition[29]. These matrices (and by implication _ (s)) appear to describe "quantum systems having classical analogues that display chaotic behaviour and are not symmetric under time-reversal", which sounds rather like the resistivity of a 2DEG immersed in a MW bath. It would be interesting to study the properties of these matrices in the presence of a magnetic field. Physicists may find random matrices and the localized states model more appealing than Denjoy's probabilistic interpretation[27] of the Riemann hypothesis.

The hysteretic or re-entrant nonlinear quantum Hall anomalies that resemble pinned or sliding charge density waves[26] occur for _ > 2 and most clearly at _ = 4_ Landau level filling at 10 Hz. The most prominent periodic correction to (3), of unexplained origin, is Chybyshev's bias (1853): there are many more primes of the form $4n + 3$ (configuration space floppy) than of the form $4n + 1$ (configuration space rigid, hence pinning). Is this a coincidence? Is there an alternative explanation for the exceptional clarity of the $\nu = 4$ anomaly? What about the Chybyshev harmonic, 8n + 1, etc.? There are no biases for 8n + 3 compared to 8n + 1 below $10^{12}$, but there is a large bias[30] near $10^{14}$. The Chybyshev harmonic at $\nu = 8$ Landau level filling is inaccessible at 10 Hz with only $10^{10}$ carriers, but it might be accessible at the harmonic average of 10 Hz and 10GHz, that is, at radio frequencies, and higher voltages.